# *"Fragile superconductivity"*: a kinetic glass transition in the vortex matter of the high-temperature superconductor YBa$_2$Cu$_3$O$_{7-\delta}$

Rolf Lortz, [1,2]♣ Christoph Meingast, [1] Alexandre I. Rykov, [3] Setsuko Tajima [3]♠

[1] *Forschungszentrum Karlsruhe, Institut für Festkörperphysik, 76021 Karlsruhe, Germany*
[2] *Fakultät für Physik, Universität Karlsruhe, 76131 Karlsruhe, Germany*
[3] *Superconductivity Research Laboratory-ISTEC, 10-13 Shinonome I-Chome, Koto-ku, Tokyo 135, Japan*

Using high-resolution thermal expansion and magnetization measurements, we provide experimental evidence for a kinetic glass transition in the vortex matter of YBa$_2$Cu$_3$O$_{7-\delta}$ with some disorder. This transition, which represents the true superconducting transition in a magnetic field, exhibits many of the features of the usual glass transition found in supercooled structural liquids such as window glass. We demonstrate, using both kinetic and thermodynamic criteria, that this vortex matter is the most fragile system known to date, which we argue makes it possible to investigate the behavior very close to the Kauzmann temperature. Vortex matter, we suggest, may be a model system to study glassy behavior in general, which is expected to lead to a better understanding of the strong-fragile behavior in structural glasses.

*Keywords: kinetic glass transition, vortex matter, thermal expansion, flux pinning, high-temperature superconductors*

## 1. INTRODUCTION

Of the many different kinds of glassy states found in nature, spin glasses and structural glasses (e.g. normal window glass) have probably received the most attention. One of the central questions concerning the glassy state is whether the 'glass' is a distinct thermodynamic phase from the 'liquid' phase[1]. It is believed that this may actually be the case for spin glasses[2], whereas the glass transition in structural glasses appears to be mostly kinetic in nature[3]. Supercooled structural liquids, however, exhibit a wide range of behaviors, which can be classified using the 'strong-fragile' scheme of Angell[3], where the terms 'strong' and 'fragile' refer to Arrhenius and highly non-Arrhenius relaxation behaviors, respectively. For the extremely fragile liquids, a thermodynamic origin of the glass transition cannot be ruled out. A microscopic understanding of the complexity leading to different degrees of fragility and affecting the kinetic glass transition is, however, lacking[1]. Also the connection, if any, between the different kind of glass states (spin, structural, etc.) is unclear. Here we examine the glassy behavior of a novel type of matter, namely the vortex matter in the high-temperature superconductor (HTSC) YBa$_2$Cu$_3$O$_x$.

## 2. MOTIVATION

Vortex matter is made up of interacting quantised vortices, which are formed when a magnetic field penetrates a type-II superconductor. The physics of vortex matter has recently attracted considerable attention because of the complex phase diagrams, which result from the

---

♣ Present address: Department od Condensed Matter Physics, University of Geneva, 24 Quai Ernest-Ansermet, CH-1211 Geneva 4, Switzerland, E-mail: Rolf.Lortz@physics.unige.ch

♠ Present address: Department of Physics, Osaka University, Machikaneyama-cho 1-1, Toyonaka-shi, Osaka 560-0043, Japan

competition of vortex-vortex, superconducting layer-layer coupling, vortex pinning, and thermal energies[4]. In very clean samples, the low-temperature crystalline vortex phase melts via a first-order phase transition[4-7]. The inset of Fig. 1 shows entropy data near this melting transition in $YBa_2Cu_3O_x$ crystals, which were obtained by Schilling *et al.* using high-resolution specific-heat measurements[8]. At $T_m$ = 84.2 K one clearly notices a jump in the entropy, illustrating the first-order nature of the melting transition. Above $T_m$, the entropy continues to rapidly increase due to the higher heat capacity of the liquid compared to that of the solid. An extrapolation of the entropy of the liquid below $T_m$ (red dotted line) shows that the entropy of the supercooled vortex liquid would fall below that of the vortex solid already about 1 K below $T_m$. This crossing point defines the Kauzmann temperature $T_K$ (We note that the extrapolation of the entropy of the vortex liquid below $T_m$ is unproblematic due to the close proximity of $T_K$ and $T_m$.), and, since the famous paper by Kauzmann in 1948[9], the question of what would happen to a supercooled liquid at this temperature has remained one of the great unresolved problems in physics/chemistry. This is partly because the kinetic glass transition, which always occurs above $T_K$, prevents one from studying the equilibrium behavior of the supercooled liquid at $T_K$. Recently, Ito *et al.*[10] showed that the ratio $T_K/T_m$ is a good indicator of the fragility (or complexity) of supercooled liquids, and in the main part of Fig. 1 we reproduce the Kauzmann-type plot of various supercooled glass forming liquids from Ref.10 and add our extrapolation of the vortex-matter entropy data from the inset. The value of $T_K/T_m$ for $YBa_2Cu_3O_x$ suggests that the vortex liquid is the most fragile liquid known to date.

The vortex-solid to vortex-liquid transition in clean $YBa_2Cu_3O_x$ crystals shows only a very small hysteresis, and it does not appear feasible to significantly supercool the vortex liquid[5,6]; the reader may thus wonder about the significance of Fig. 1 and the above discussion about $T_K$ in the vortex matter in HTSCs. With the introduction of random vortex pinning defects, however, the first-order melting transition vanishes and is replaced by a more continuous transition[4,11-14], commonly referred to as the 'vortex-glass' transition. There is evidence for different glassy phases (Bose glass, Bragg Glass and vortex glass)[4], the detailed nature of which however remains controversial. There are experimental and theoretical indications that the vortex-glass transition is in fact a true second-order phase transition with critical behavior similar spin glasses[15]. In contrast, some authors suggest a kinetic glass transition, similar to the usual structural glass transition in supercooled liquids[16]. In the following, we present experimental evidence for the latter picture, a kinetic glass (not a thermodynamic) transition with the same high fragility as suggested by the thermodynamic data of Fig. 1.

### 3. EXPERIMENTS

High-resolution thermal expansion and magnetization measurements have been performed on a $YBa_2Cu_3O_{7.0}$ single crystal (with linear dimensions of 6 x 3 x 2 $mm^3$ for the different orthorhombic a, b and c axes) grown by a pulling technique, detwinned at a uniaxial pressure of 10 MPa and annealed at 376 bar oxygen pressure and 400°C for 140 hours. These crystals do not exhibit reversible properties in magnetic fields, probably due to random impurity pinning centers. The thermal expansion was measured in a high-resolution capacitance dilatometer in the range 7-140 K in magnetic fields up to 11.4 T. The magnetization data were taken with a vibrating sample magnetometer in a constant field of 6 T. Both types of measurements were performed in a continuous cooling and heating mode with rates between ±2 and ±90 mK/s. These types of temperature scanning experiments are ideally suited to probe the kinetics of glass transitions, since the time scale of the experiment is directly proportional to the heating (cooling) rate[17-21].

In Fig. 2 a) we present the thermal expansivity, $a(T) = (1/L) \cdot (dL/dT)$ ($L$ is the length of the sample), for the orthorhombic b-axis in magnetic fields up to 11.4 T applied parallel to the c-axis. In addition to the superconducting anomaly at the $H_{C2}(T)$-crossover (appearing as a jump at zero field, which broadens with increasing field), surprisingly large and sharp peaks appear in the data above 3.4 T and at temperatures $T_g$ slightly below what would be the vortex melting temperature in clean crystals. These anomalies are not of purely thermodynamic origin, as the history dependence of the heating curves at 6.8 T in Fig. 2b clearly demonstrate. The data upon heating at the same rate (18 mK/s) depend strongly on the previous cooling cycle, and an anomaly is hardly detectable upon cooling. Moreover, the large positive peaks in Fig. 2a, which were all measured upon heating at a rate of 18 mK/s after cooling with 90 mK/s, can even be made negative if the cooling rate is made smaller than the heating rate (see Fig. 2b). Additionally, the anomaly shifts to higher temperatures for higher heating rates after the same fast cooling (-90 mK/s) (see inset in Fig. 2) (These small temperature shifts could be accurately determined only because the anomaly at $T_c$ provided a close calibration point.). Both the history-dependent shape of the anomaly and the heating-rate dependence are typical for kinetic glass transitions[17-21]. We note that the large peaks in $a(T)$ most likely are caused by magnetostrictive effects[22] as a result of macroscopic currents, which decay upon heating at the glass transition. The rate of decay of these currents is governed by the movement of the vortices, and, thus, provides a measure of the relaxation of the vortex matter. In contrast to usual glass transitions, little effect is seen upon cooling at $T_g$, because these currents slowly build up below $T_g$.

The glasslike falling out of equilibrium can also be observed in the magnetization data upon cooling and heating in a constant field (see Fig. 2 c and also Ref. 23). The diamagnetic moment, which increases upon cooling below $T_c$, freezes at the glass transition temperature ($T_g$), resulting in a relatively sharp kink. The faster the crystal is cooled, the higher is the temperature where the kink occurs and the lower is the diamagnetic signal. Upon heating after fast cooling, the diamagnetic signal shows a significant hysteresis and relaxes towards the equilibrium state at the same temperature where the expansivity shows a peak (see Fig. 2c).

## 4. DISCUSSION

The simplest thermally activated behavior, where the relaxation follows an Arrhenius law

$$\tau(T) = \nu_0^{-1} \cdot e^{\frac{E_a}{k_B T}}, \qquad (1)$$

is found for strong glasses where individual relaxation processes are independent[1,3]. $\tau$ at $T_g$ is directly related to the heating rate, r, in our experiment by[20]:

$$\tau(T_g) = \frac{k_B \cdot T_g}{E_a} \cdot \frac{T_g}{|r|}. \qquad (2)$$

Although the inset in Fig. 2 shows $\tau(T_g)$ for various heating rates in an Arrhenius plot with a good linear dependence, the extracted activation energy $E_a$=2.4±0.1 eV and attempt frequency $\nu_0$=$10^{366\pm50}$ Hz are unphysical. A value of $E_a$ on the order of 1 eV is typical for the potential barrier for rearrangements of atoms in a solid, but much lower energies are expected for the vortex-pinning interaction. Such 'unphysical' relaxation parameters are, however, observed for highly complex relaxation phenomena, as e.g. found in spin glasses[2] or highly fragile structural glasses[1,3] and are directly related to the highly non-Arrhenius behavior, which can be modeled as Vogel-Tamman-Fulcher behavior by replacing the exponent in Eq. (1) by $E_a/k_B(T-T_0)$. Our data thus indicate extreme fragility. A more quantitative fragility measure, which is closely related to the kinetic parameters, is given by the relative width $\Delta T_g$ of the anomaly at $T_g$. We find $\Delta T_g/T_g$ = 0.01, which suggests an extremely fragile system[10], fully consistent with the above kinetic parameters. To better demonstrate this point, we reproduce in Fig. 3 the plot of D$T_g/T_g$ versus the fragility metric $F_{1/2}$ from Ref. 3. $F_{1/2}$ in this plot is either the kinetic fragility (see Ref. [10]) or the thermodynamically derived fragility ($F_{1/2\,cal}$=2*$T_{1/2}/T_m$-1, where $T_{1/2}$ is the temperature where D$S$/D$S_m$=0.5) and varies between zero (strong) and one (fragile). As can be seen from Fig 3, the single datum for the vortex liquid from the present work fits nicely into the general correlation between the fragility metrics D$T_g/T_g$ and $F_{1/2}$ and extends the data set at the extreme fragility end. Note that we have also added a data point in Fig. 3 for the orientational glass transition in $C_{60}$, which represents a very simple system at the extremely strong limit[18,19].

The above thermodynamic and kinetic analysis of the vortex matter in $YBa_2Cu_3O_{7.0}$ suggests that the presently observed kinetic glass transition follows the same phenomenology as found in supercooled structural liquids despite the fact that the vortex system is not supercooled in the traditional sense. In the vortex system, the first-order crystallization is prevented not by supercooling, but rather through pinning centers, which can effectively slow down the dynamics of vortex equilibration, however, without significantly disturbing the underlying thermodynamics[24]. Thus, although a small amount of pinning centers prevent crystallization of the vortices, the step-like increase in the specific heat shown in the liquid phase (see inset of Fig. 1) is a robust feature, which is unaffected by small densities of pinning centers[25], and it is just this feature which is most likely responsible for the high fragility of the vortex matter. In terms of the Kauzmann diagram, fragility is defined by the slope of the liquid entropy vs. $T$ curve; the higher the slope, the higher the fragility. This slope is given by the ratio specific heat jump divided by the melting entropy D$S_m$. D$S_m$ for the vortex matter of $YBa_2Cu_3O_{7.0}$ has a very usual value in terms of entropy per 'particle', which suggests that cause of the fragility must lie in the value D$C_p$, which in fact is anomalously large[7]. The origin of this large D$C_p$ are unknown, but may be linked to the strong thermal phase fluctuations of the superconducting order parameter in the liquid phase[26], which can be visualized in terms of vortex-loop excitations[26,27] and may directly affect the kinetic glass transition through entanglement of vortices. In structural glasses there is the possibility of a liquid-liquid phase transition coming in at extreme fragilities, which may contract to a critical point under certain circumstances[29,30]. The vicinity of such a critical point to $T_g$ could provide another explanation for the extreme heat capacity of the vortex glass former. Indeed, interesting features have been observed recently in the vortex matter phase diagram of high-temperature superconductors, such as inverse melting below a Kauzmann point[24,28] and evidence for a liquid-liquid transition in its vicinity[14]. These might be related to the features observed in structural glasses.

## 5. CONCLUSIONS

In conclusion, our results provide evidence for a highly fragile kinetic transition and not a true thermodynamic transition in the vortex matter of $YBa_2Cu_3O_{7.0}$ with some pinning. The first-order vortex melting transition in clean systems is generally regarded as the true superconducting transition in a magnetic field[4,5], and our results, thus, suggest that the true superconducting transition in the presence of disorder is in fact a kinetic and not a true thermodynamic transition. The high fragility suggests that this glass transition is not solely due to disordering of vortices and that additional degrees of freedom, such as phase fluctuations or vortex-loop excitations, must be considered as well. The close similarities with structural liquids and the easy tuneability of the vortex system through field and disorder suggest that vortex matter may be a model system to study melting, glass formation and fragility in general[31].

**ACKNOWLEDGMENTS**

We would like to thank W. Goldacker for use of his magnet system and H. Küpfer for making the magnetization measurements. We also acknowledge enlightening discussions with H. v. Löhneysen, P. Nagel, M. E. Reeves and S. Schuppler.

**Figure Captions**

**Fig. 1.** Inset: Excess entropy $\Delta S$ of the vortex matter in $YBa_2Cu_3O_{7-\delta}$ taken from Ref. [8], where the vertical arrow indicates the entropy jump $\Delta S_m$ at the melting transition. The red dotted line is an extrapolation of the entropy of the liquid below $T_m$ to the Kauzmann temperature $T_K$, where the entropy of the liquid would fall below that of the crystalline phase. Main figure: Kauzmann plot ($\Delta S/\Delta S_m$ versus $T/T_m$) with data of various supercooled glass forming liquids taken from Ref. [10]. The extrapolation of the entropy of the vortex liquid to values below the vortex lattice melting temperature suggests that the vortex liquid is extremely fragile.

**Fig. 2. a)** Linear thermal expansivity data for the orthorhombic b-axis of a $YBa_2Cu_3O_{7.0}$ single crystal in magnetic fields from 0 to 11.4 T applied parallel to the c-axis. The data were all taken upon heating ( +18 mK/s) after cooling at -90 mK/s. b) History dependence of the thermal expansivity anomaly at $H$=6.8 T upon heating (+18 mK/s) after cooling at various rates between -10 mK/s and -90 mK/s (Only the -10 mK/s cooling curve is shown.). Inset: Expansivity anomaly for heating rates between +2 and +95 mK/s (cooling rate = -90 mK/s), as well as an Arrhenius plot of the relaxation rate derived from temperature shift of these data (see text for details). c) Magnetization data taken at different cooling and heating rates in a constant field of 6 T. The magnetization falls out of equilibrium at the same temperature at which the anomaly in the expansivity occurs (blue dotted line).

**Fig. 3.** Correlation between fragility metrics $\Delta T_g/T_g$ and $F_{1/2}$ for various supercooled liquids taken from Ref. [10], for $C_{60}$ [18], and for the vortex liquid (red datum). Both $\Delta T_g/T_g$, which was taken from our thermal expansion data, and the thermodynamic $F_{1/2\ cal} = (2*(T_{1/2}/T_m)-1)$, which was taken from Fig. 1, suggest that the vortex liquid is the most fragile glassy system known to date.

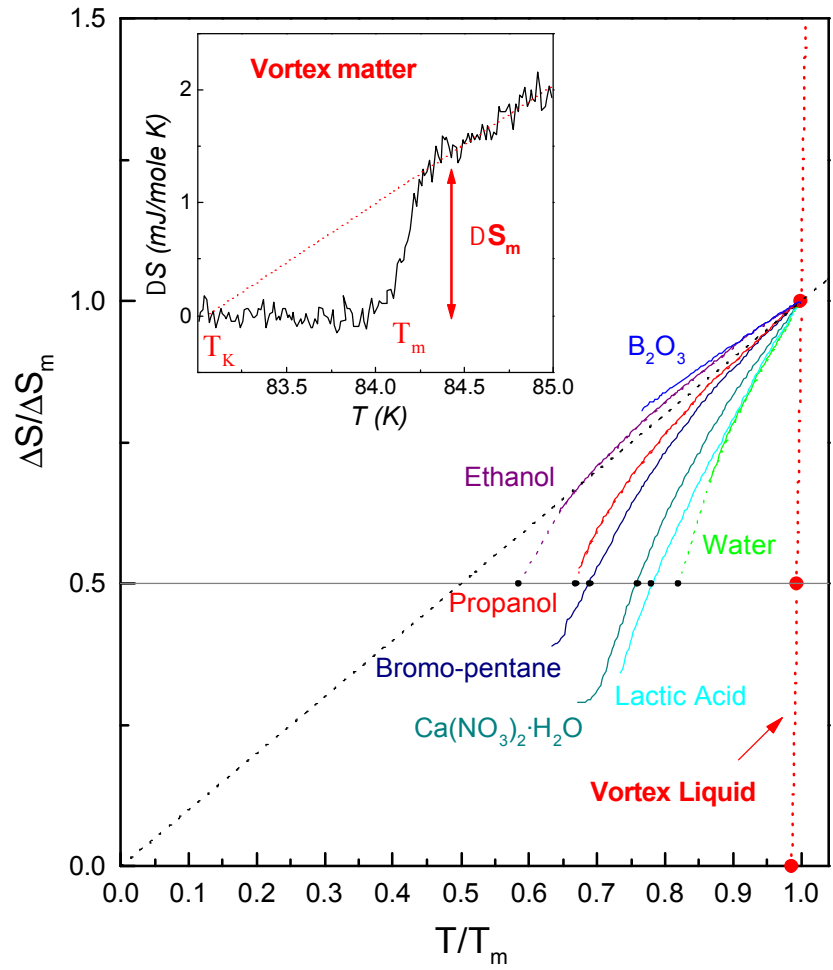

**Fig. 1.**

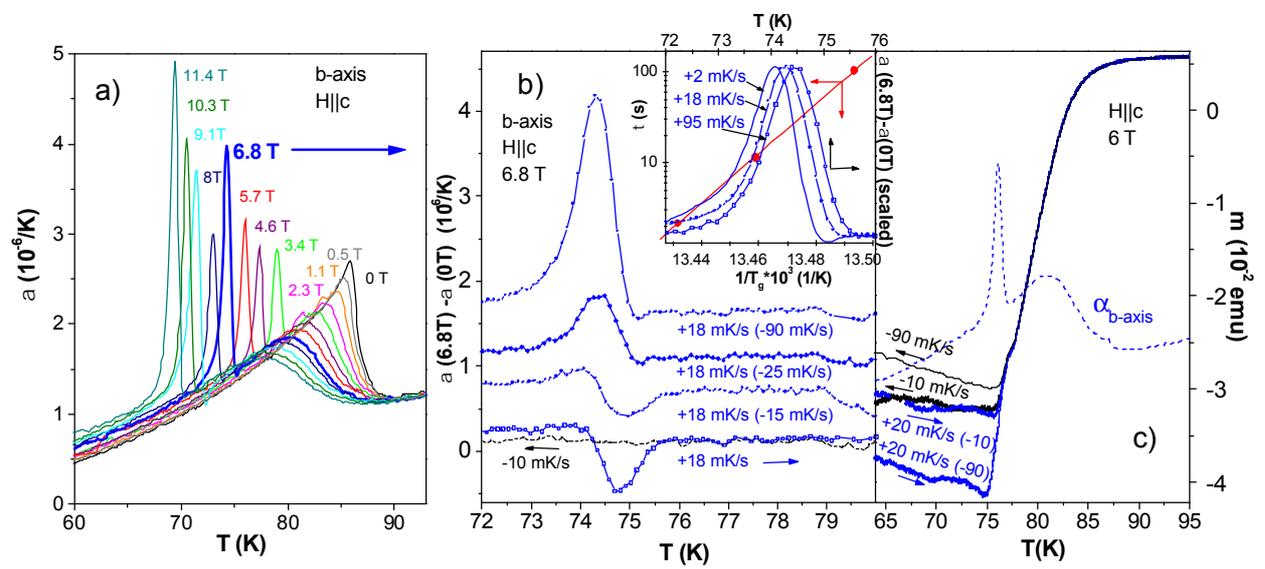

**Fig. 2a**
**Fig. 2b**
**Fig. 2c**

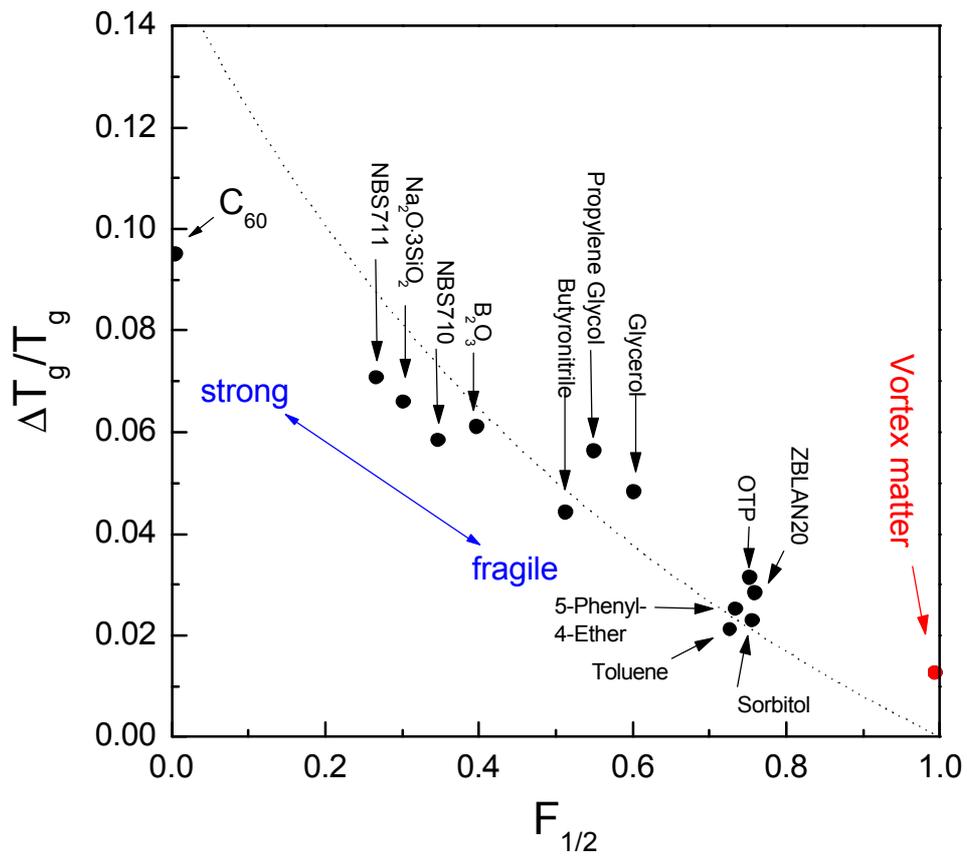

**Fig. 3.**